# Deformation-Aware Segmentation Network Robust to Motion Artifacts for Brain Tissue Segmentation using Disentanglement Learning

Sunyoung Jung[1], Yoonseok Choi[1], Mohammed A. Al-masni[2], Minyoung Jung[3] & Dong-Hyun Kim[1]

[1] Department of Electrical and Electronic Engineering, College of Engineering, Yonsei University, Seoul, Republic of Korea
`{sunyoungj, yoonseokchoi, donghyunkim}@yonsei.ac.kr`

[2] Department of Artificial Intelligence and Data Science, College of Software and Convergence Technology, Sejong University, Seoul, Republic of Korea
`m.almasani@sejong.ac.kr`

[3] Cognitive Science Research Group, Korea Brain Research Institute, Daegu, Republic of Korea
`minyoung@kbri.re.kr`

**Abstract.** Motion artifacts caused by prolonged acquisition time are a significant challenge in Magnetic Resonance Imaging (MRI), hindering accurate tissue segmentation. These artifacts appear as blurred images that mimic tissue-like appearances, making segmentation difficult. This study proposes a novel deep learning framework that demonstrates superior performance in both motion correction and robust brain tissue segmentation in the presence of artifacts. The core concept lies in a complementary process: a disentanglement learning network progressively removes artifacts, leading to cleaner images and consequently, more accurate segmentation by a jointly trained motion estimation and segmentation network. This network generates three outputs: a motion-corrected image, a motion deformation map that identifies artifact-affected regions, and a brain tissue segmentation mask. This deformation serves as a guidance mechanism for the disentanglement process, aiding the model in recovering lost information or removing artificial structures introduced by the artifacts. Extensive in-vivo experiments on pediatric motion data demonstrate that our proposed framework outperforms state-of-the-art methods in segmenting motion-corrupted MRI scans. The code is available at https://github.com/SunYJ-hxppy/Multi-Net.

## 1  Introduction

Brain tissue segmentation plays a substantial role in medical image analysis. It serves to accurately identify and visualize important anatomical structures, enabling the diagnosis of diseases, especially neurodegenerative disorders such as Alzheimer's and

Deformation-Aware Segmentation Network Robust to Motion Artifacts

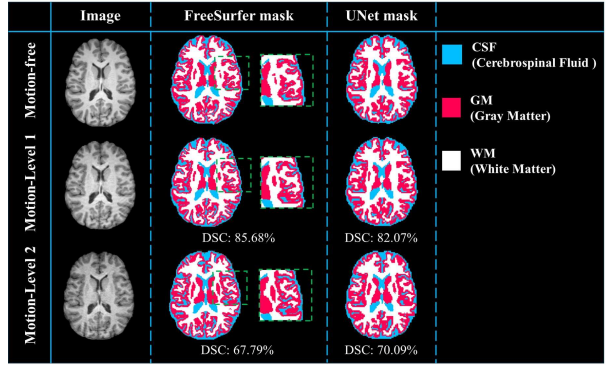

**Fig. 1.** Segmentation masks generated by FreeSurfer [1] and U-Net on motion-free and motion-corrupted in-vivo data at two severity levels. Compared to clean data, motion artifacts cause blurring and structural variations, leading to imprecise segmentation masks. The Dice similarity coefficient (DSC) represents the average DSC within the White Matter (WM), Gray Matter (GM), and Cerebrospinal Fluid (CSF).

Parkinson's diseases [2]. It is also essential for studies on early infant brain development and quantitative tissue assessments [3]. Deep learning, particularly Convolutional Neural Networks (CNNs), has emerged as a powerful tool for brain tissue segmentation [4].

However, these segmentation methods encounter challenges when handling Magnetic Resonance (MR) images affected by motion artifacts, which are prevalent in clinical setting due to the prolonged data acquisition duration of MRI scans. This issue is further exacerbated in pediatric patients who struggle to remain still, resulting in increased motion artifacts during scans [5, 6]. As shown in **Fig. 1**, motion artifacts degrade image quality by the presence of ringing, blurring, and ghosting artifacts [7]. Consequently, this significantly hinders the precise analysis of MR image, particularly in estimating cortical gray matter volume and thickness when employing traditional segmentation methods [8]. Notably, the severity of motion artifacts directly correlates with the degree of degradation in segmentation performance.

Addressing these artifacts in the volumetric data often requires additional MRI scans, leading to prolonged acquisition time, cost, and inconvenience for subjects. Even slight movements with a few millimeters can result in erroneous segmentation predictions [9]. Therefore, developing robust segmentation methods that handle motion-corrupted data remains a critical challenge.

In this context, previous studies have addressed motion correction and segmentation, primarily focusing on non-rigid motion in cardiac [10, 11] or lung images [12]. Additionally, a study examined the impact of motion-artifacts on the quality of cortical reconstruction in brain MR images [13]. Researchers demonstrated that reconstructing the brain cortex after motion correction improves the quality of the reconstruction. However, this was not an end-to-end model; instead, segmentation was used solely to estimate their proposed motion-correction network. Furthermore, research has investigated segmentation methods that are robust to motion artifacts by utilizing

Deformation-Aware Segmentation Network Robust to Motion Artifacts

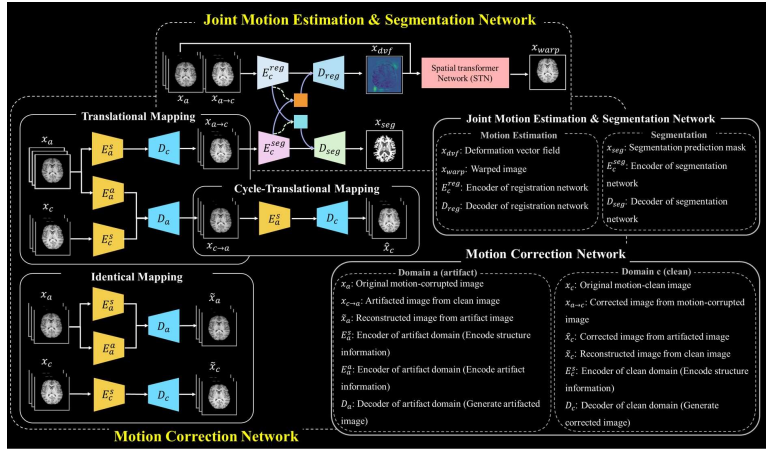

**Fig. 2.** Overview framework of the proposed network.

several types of images, including images affected by motion, those without motion, and images with different resolution [14, 15]. These approaches prioritized the utilization of large datasets rather than the development of new methodologies.

In this paper, we propose a robust segmentation framework for motion artifacts, leveraging the prior knowledge of motion obtained through a jointly stitched motion estimation network. The proposed framework comprises three interconnected components: a disentanglement learning network, a segmentation network, and a motion estimation network. Our work presents four significant contributions. First, we develop a novel self-complementary network with a 2.5D disentangled dual-domain design. Second, we propose a motion-aware segmentation network by integrating it with a jointly stitched motion estimation network. Third, a motion deformation map is employed to maintain consistency between the disentangled artifact style and the anatomical structure of the brain. Finally, we validate the proposed method using in-vivo pediatric motion data.

## 2 Methods

Our proposed model is trained in an end-to-end manner, integrating a motion correction disentanglement learning network with a joint motion estimation and segmentation network as illustrated in **Fig. 2**.

For motion correction, the disentanglement learning separates motion artifacts from the underlying anatomical structure within the motion-corrupted image. This corrected image is then fed into the jointly stitched motion estimation and segmentation network, which leverages crucial artifact features for generating an accurate tissue segmentation mask.

Given that our model is designed to perform three tasks simultaneously, we have chosen to use a 2.5D method to address challenges related to model complexity and resource requirements.

## 2.1 Data Acquisition and Preprocessing

In this study, we utilized 3D T1-weighted brain MRI scans from two sources: a private dataset containing unpaired pediatric scans, and a public dataset from OpenNeuro [16], which provided three types of paired data: motion-clean scans and two levels of motion-corrupted scans. We used 2,847 axial slices for training and 608 axial slices for evaluation. Additionally, we included 955 slices of 2D motion-corrupted images from pediatric in-vivo motion artifact experiments.

The public dataset was acquired using a 3T MRI scanner (MAGNETOM Prisma, SIEMENS, Germany) with the following parameters: echo time (TE) of 3ms, repetition time (TR) of 2300ms, flip angle of 9°, and field of view (FOV) of 256 × 256 [mm]. The private dataset was acquired using a 3T MRI scanner (MAGNETOM Skyra, SIEMENS, Germany) with the following parameters: TE of 2.3ms, TR of 2400ms, a flip angle of 8°, and FOV of 230 × 230 [mm]. The axial image resolution was 1 × 1 [mm] for the public dataset and 0.7 × 0.7 [mm] for the private dataset. To achieve consistent resolution in both datasets, we applied preprocessing techniques to standardize the image size to 256 × 256 pixels.

To generate datasets affected by motion, we utilized a motion simulator called View2DMotion [17]. These datasets were derived from a private collection that did not have paired motion-clean and motion-corrupted data. To utilize the entire brain volume as a 2.5D input, we extracted brain slices in an axial view and concatenated every three adjacent slices together. Subsequently, the datasets were resized to a uniform size of 3 × 256 × 256.

## 2.2 Disentanglement Network

The disentanglement learning network is inspired from the UDDN architecture [18]. Further details regarding its implementation are provided in the following sections.

**Translational Mapping.** Translational mapping process is the key component within the motion correction network. It aids in learning the invertible mapping between motion-corrupted and clean distributions during the training stage and is used to remove artifacts during the testing stage. For motion-corrupted image, two encoders were employed: $E_a^s(.)$ and $E_a^a(.)$ were designed to extract structural and artifact components, respectively. For the clean image, $E_c^s(.)$ was utilized, which was designed to extract only structural components.

$$s_a = E_a^s(x_a), a = E_a^a(x_a), s_c = E_c^s(x_c) \qquad (1)$$

After extracting the components, the artifact component $a$ was switched from being with the structural components $s_a$ of the artifact image to align with the structural components $s_c$ of the clean image. Following this, two decoders were introduced to generate new images, the motion corrupted image $(x_{c \to a})$ and clean image $(x_{a \to c})$. The clean images were reconstructed using only the structural component $s_a$ derived from artifact image. The artifact images were then reconstructed utilizing the structural

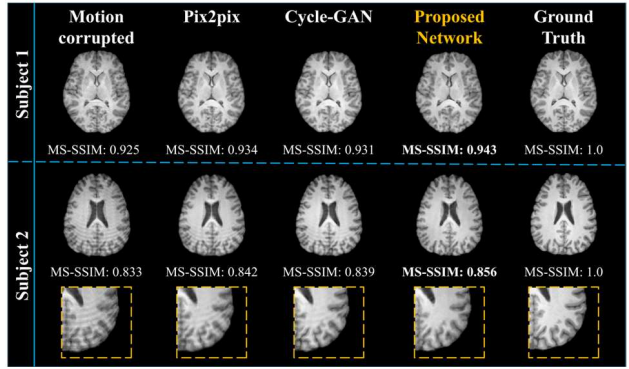

**Fig. 3.** Results of motion correction using OpenNeuro data with in-vivo motion.

component $s_c$ from the clean image and the artifact component a from the artifact image.

$$x_{a \to c} = D_a(s_a), x_{c \to a} = D_c(s_c, a) \tag{2}$$

**Cycle-Translational Mapping.** In cycle-translational mapping process, we redo the disentanglement, component switch, and image generation with the output of the translational mapping ($x_{c \to a}$). We added cycle translational mapping to ensure consistency between images and stabilize the generation process. However, to reduce the computational cost and learning time, we performed the cycle translational mapping once, from the generated motion-corrupted image back to the clean image.

The method of encoding and decoding is the same as the method used in the translational mapping.

$$\hat{s}_a = E_a^s(x_{c \to a}), \hat{a} = E_a^a(x_{c \to a}) \tag{3}$$

$$\hat{x}_c = D_c(\hat{s}_a) \tag{4}$$

**Identical Mapping.** Identical mapping serves as a reconstructive approach. By simply utilizing the components for generation without any interchange, it aids in preserving the original structure and content of the image. This process helps to assist in motivating the network to preserve the unaltered quality of the image.

$$\tilde{s}_a = E_a^s(x_a), \tilde{a} = E_a^a(x_a), \tilde{s}_c = E_c^s(x_c) \tag{5}$$

$$\tilde{x}_a = D_a(\tilde{s}_a, a), \tilde{x}_c = D_c(\tilde{s}_c) \tag{6}$$

### 2.3 Deformation-Aware Segmentation Network

The deformation-aware segmentation network comprises an integrated framework of motion estimation and segmentation networks.

**Motion Estimation Network.** For the motion estimation network, we derived our concept from the registration network [19]. Two datasets, labeled as source and target

Deformation-Aware Segmentation Network Robust to Motion Artifacts

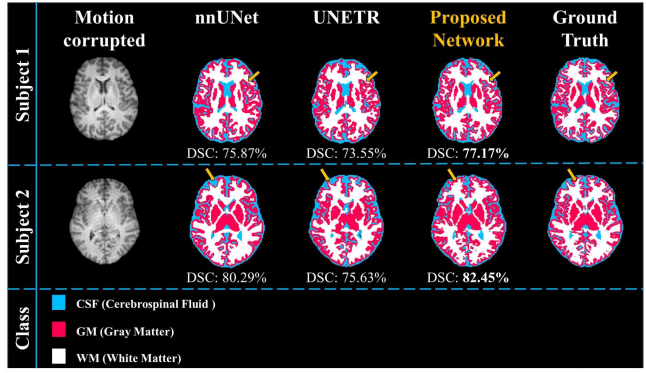

**Fig. 4.** Results of tissue segmentation using OpenNeuro data. The value represents the average DSC of GM and WM.

are concatenated to serve as the input of the registration network. In our proposed method, registration model is utilized as a motion correction network, where motion-corrupted data ($x_a$) is defined as source data, and motion-corrected data ($x_{c \to a}$) from the translational mapping as target data.

Given that the network takes a pair of images as input, it learns the image features that captures the information of the Deformation Vector Field (DVF), representing displacements between a paired source and target images. Subsequently, the DVF and the source data are processed through the Spatial Transformer Network (STN) and warped to match the target data. In our network, we assume that the DVF exhibits a similar form to the motion artifact.

$$F_{reg} = E_c^{reg}(x_a, x_{c \to a}), x_{dvf} = D_{reg}(F_{reg}) \quad (7)$$

$$x_{warp} = STN(x_a, x_{dvf}) \quad (8)$$

**Segmentation Network.** For the brain tissue segmentation, we utilized an optimized nnUNet [20] as our backbone network for segmenting tissue into 4 classes: (0) background (BG), (1) cerebro-spinal fluid (CSF), (2) gray matter (GM), and (3) white matter (WM). However, the segmentation network itself can result in inaccurate estimations of brain structure due to the presence of motion-corrupted MRI [21].

To develop a segmentation network that is robust to motion artifacts we incorporated an additional motion estimation network. This network aims to identify and differentiate motion artifacts during the segmentation process. By leveraging the concept of warping, we could generate motion deformation maps that help in correction of motion-corrupted images. This strategy not only offers insights into the extent of motion but also directs focus towards the impacted regions [22].

To improve the segmentation performance by utilizing information from the motion estimation network, we adopted a Multi-Task Learning (MTL) approach [23]. This approach enables the networks to share critical information such as the magnitude values of motion artifacts and the affected regions. During training, the segmentation network learns to create the segmentation mask by exploiting the motion information. To amplify the efficacy of MTL, we integrate cross-stitch units. These units

Deformation-Aware Segmentation Network Robust to Motion Artifacts

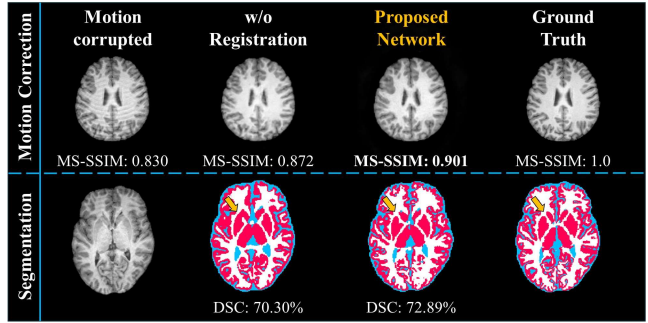

**Fig. 5.** Ablation study on different network framework. Our network is compared with a version without registration network. The value represents the average DSC of GM and WM.

dynamically adjust the weights between the feature maps from each network during training, determining their significance and enhancing the inter-network correlation [24].

## 3   Loss functions

The adversarial loss functions are used to distinguish between real and generated images in the domain of translational mapping. The mathematical formulation is as follows:

$$x_{a \to c} = D_a(s_a), x_{c \to a} = D_c(s_c, a) \tag{9}$$

$$\mathcal{L}_{adv}(Dis) = \frac{1}{2}\mathbb{E}_{x_a \sim I_a}\left[\left(1 - Dis_a(x_a)\right)^2 + \left(Dis_a(x_{c \to a})\right)^2\right]$$
$$+ \frac{1}{2}\mathbb{E}_{x_c \sim I_c}\left[\left(1 - Dis_c(x_c)\right)^2 + \left(Dis_c(x_{a \to c})\right)^2\right] \tag{10}$$

$$\mathcal{L}_{adv}(E, D) = \frac{1}{2}\mathbb{E}_{x_a \sim I_a}\left(1 - Dis_a(x_{c \to a})\right)^2 + \frac{1}{2}\mathbb{E}_{x_c \sim I_c}\left(1 - Dis_c(x_{a \to c})\right)^2 \tag{11}$$

The term $Dis(\cdot)$ refers to the discriminator, which serves to differentiate between $(x_a, x_{c \to a})$ and $(x_c, x_{a \to c})$ respectively. $\mathbb{E}_{x_a \sim I_a}$ and $\mathbb{E}_{x_c \sim I_c}$ denote the operations of expectation on the distributions $I_a$ and $I_c$, respectively. In terms of cycle consistency loss, we employed the mean absolute error (MAE) to quantify pixel-level loss and Multi-Scale Structural Similarity (MS-SSIM) index to measure structural level loss.

$$\mathcal{L}_{cyc} = \mathbb{E}_{x_c \sim I_c}\|x_c - \hat{x}_c\|_1 + \mathbb{E}_{x_a \sim I_a}\|x_a - \hat{x}_a\|_1 \tag{12}$$

In the case of identical mapping, the output is the same as the input due to the absence of component exchange. The associated loss can be calculated as follows:

$$\mathcal{L}_{idt} = \mathbb{E}_{x_c \sim I_c}\|x_c - \tilde{x}_c\|_1 + \mathbb{E}_{x_a \sim I_a}\|x_a - \tilde{x}_a\|_1 \tag{13}$$

Deformation-Aware Segmentation Network Robust to Motion Artifacts

Additional generation loss was addressed through an artifact loss, denoted as $\mathcal{L}_{art}$. We discovered that the generated correction output ($\hat{x}_c$) might seem plausible, but it may not precisely mimic the original image ($x_c$) which contains the exact structural information. To counteract this issue, we devised a new loss function, $\mathcal{L}_{art}$. This was accomplished by computing the difference between the deformation map and the precise artifact shape, which is derived from subtracting the original input images $x_c$ and $x_a$.

$$\mathcal{L}_{art} = \mathbb{E}_{x_a \sim I_a}\left[\left\|(x_a - x_c) - x_{def}\right\|_1\right] \tag{14}$$

This loss is predicated on the observation that the disparity between $x_a$ and $x_c$ should bear resemblance to the disparity between $x_a$ and $\hat{x}_c$. Because of the definition of deformation map, this disparity is the difference between the input images of the motion estimation network ($x_a$, $\hat{x}_c$).

**Table 1.** The quantitative assessment of segmentation and motion correction models consists of utilizing test datasets, which comprise both private and public datasets.

| Region | Evaluation of segmentation methods (DSC [%]) | | |
|---|---|---|---|
|  | nnUNet | UNETR | **Proposed** |
| CSF | 57.66 ± 0.14 | 54.32 ± 0.12 | **59.13 ± 0.10** |
| GM | 67.31 ± 0.13 | 64.25 ± 0.10 | **68.34 ± 0.10** |
| WM | 79.84 ± 0.09 | 77.50 ± 0.08 | **80.25 ± 0.06** |
| Metrics | Evaluation of motion correction methods | | |
|  | Corrupted motion | Pix2pix | Cycle-GAN | **Proposed** |
| MS-SSIM | 0.810 ± 0.05 | 0.875 ± 0.10 | 0.907 ± 0.04 | **0.918 ± 0.05** |
| PSNR | 23.02 ± 3.06 | 22.71 ± 3.65 | 23.48 ± 2.60 | **26.30 ± 4.17** |
| MSE | 0.010 ± 0.019 | 0.009 ± 0.014 | **0.005 ± 0.003** | 0.006 ± 0.003 |

For the motion estimation network, we employed L1 loss between $x_c$ and $x_{warp}$. By contrasting the difference between the original clean image and the output of STN, we succeeded in training the registration network. Moreover, while $x_{warp}$ gradually becomes similar to $x_{c \rightarrow a}$, it fundamentally provides constraint on the disentanglement learning process.

## 4      Results and Discussion

**Fig. 3** and **Fig. 4** illustrate a comparison of our proposed method against alternative models which are utilized for individual tasks. Highlighted by the yellow arrows in **Fig. 4,** our proposed model is observed to precisely trace the boundaries of each brain tissue, classifying them with heightened accuracy. As noted in the introduction, estimating cortical thinning across the frontal and temporal cortex is crucial in diagnosing

Parkinson's Disease [25]. **Fig. 5**, which includes an ablation study, provides evidence of the effectiveness of the registration network that was stitched together. It demonstrates that the registration network assisted the segmentation network in accurately identifying and segmenting the regions affected by the artifact.

**Table 1** presents the quantitative performance scores of our proposed model for both segmentation and motion correction on the in-vivo motion-corrupted images. The efficacy of including the artifact loss ($\mathcal{L}_{art}$) and the complementary role of the segmentation network are evident when compared to the base UDDN model, used solely for motion correction. The proposed artifact loss helps to effectively disentangles artifacts from structures, facilitating the generation of detailed motion-corrected images through the application of additional constraints. These refined images, subsequently utilized by our segmentation network, have yielded promising results.

As per the results presented in **Table 1**, our proposed network outperforms other methods in terms of motion correction. However, the segmentation results were imperfect, which can be attributed to the low resolution of the generated motion-corrected images compared to the original brain MR image. Therefore, future research should focus on generating high-resolution motion-corrected images to enhance the performance of the segmentation network.

## 5 Conclusion

In this paper, we proposed a novel motion correction and deformation aware segmentation network to generate motion corrected image and predict accurate brain tissue segmentation. Our method was evaluated on in-vivo motion-corrupted data and achieved superiority in motion correction and segmentation.

**Acknowledgement.** This work was supported by the National Research Foundation of Korea (NRF) funded by the Korean government (MSIT) (No. RS-2023-00243034).

**Disclosure of Interests.** The authors have no competing interests to declare that are relevant to the content of this article.